# SOME FURTHER COMMENTS

The author gratefully acknowledges an anonymous correspondent who politely pointed out that the caption in Figure 2 should be changed from "the integral... cannot be explicitly solved" to "... cannot be explicitly solved *by idiots.*" The integral in fact *is* solvable by non-idiots using the well known technique of scanning the pages of Gradshteyn and Ryzhik [Table of Integrals, Series, and Products, Academic Press, 1965]; reference to equation 3.386 in that text leads to the exact closed form solution:

$$I(\alpha) := \int_0^\infty \sin^2 \pi u \; u^{-\alpha} \; du = \frac{(2\pi)^\alpha}{8\Gamma(\alpha)\sin\frac{\pi}{2}(\alpha-1)}$$

Although the basic conclusions of the paper are unaffected, a number of the expressions can be considerably simplifed by substituing this formula. This idiot apologizes to any readers who may be inconvenienced by the oversight.

Also, the derivation in section 7.1 that the $K_2$ entropy is zero agrees with the numerical observation in a recent paper by A. Provenzale, A. R. Osborne, and R. Soj ["Convergence of the $K_2$ entropy for random noises with power law spectra", *Physica D* **47** (1991) 361–372].

jt

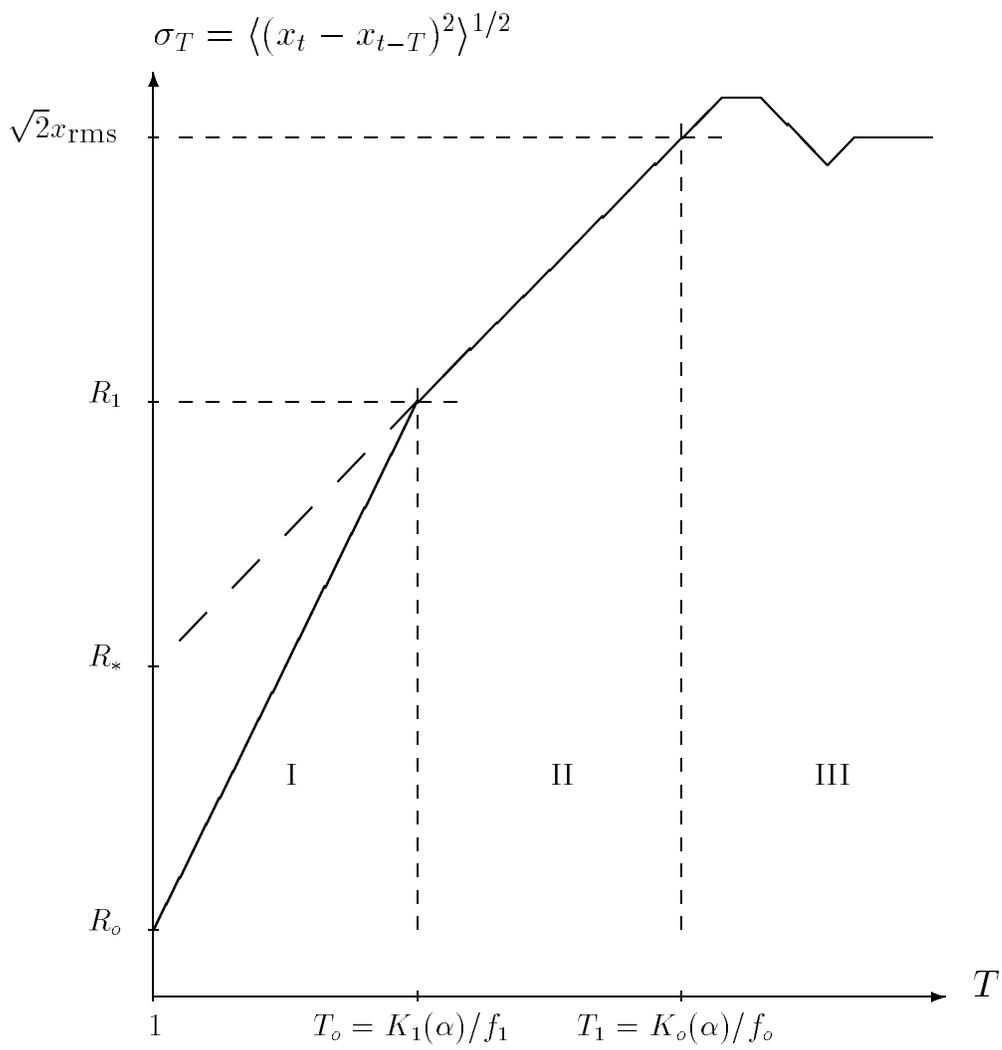

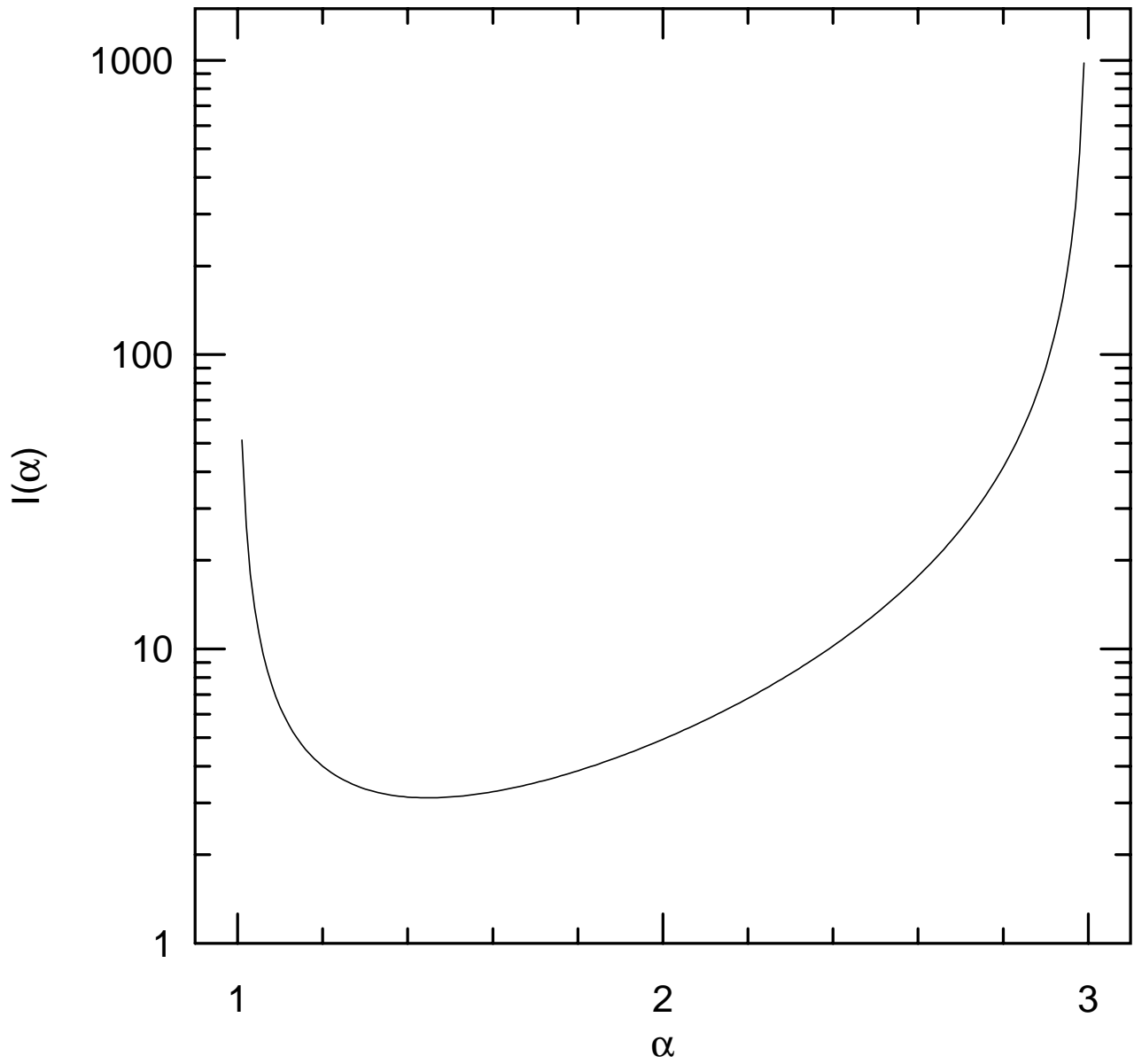

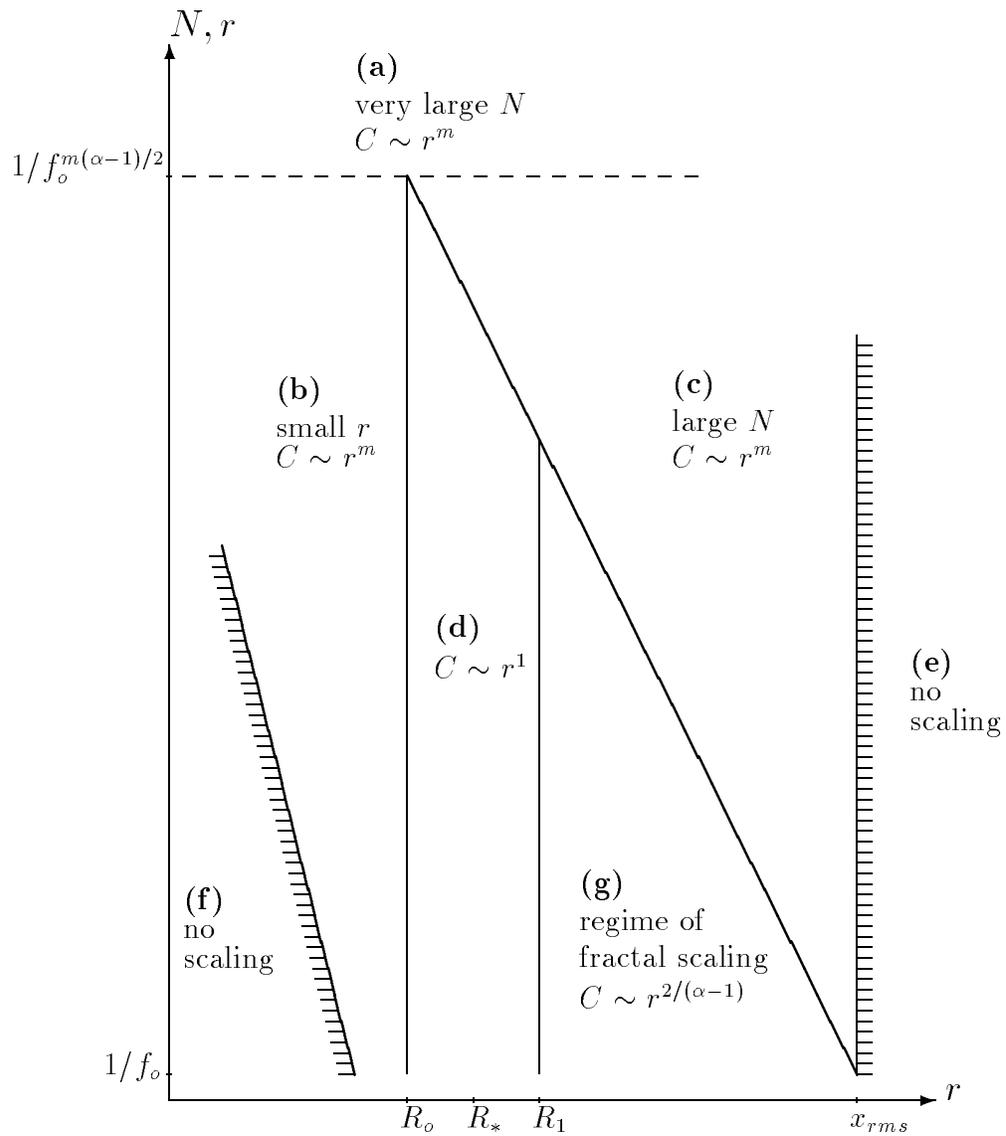

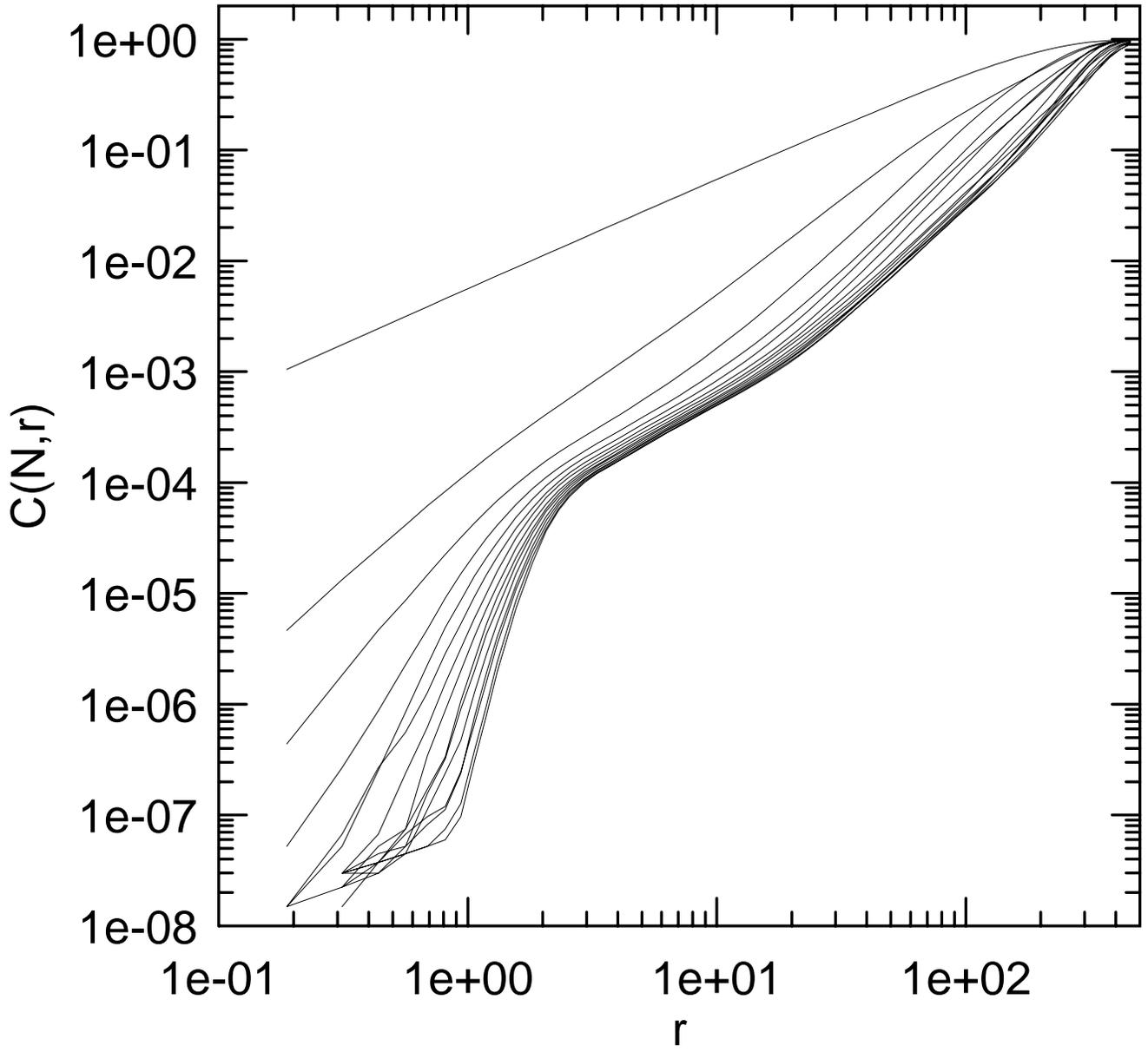

# Some comments on the correlation dimension of $1/f^\alpha$ noise

James Theiler[1]

Theoretical Division, Los Alamos National Laboratory
*Complex Systems Group, MS-B213*
*Los Alamos, NM 87545*

January 25, 1991

### Abstract

It has recently been observed that a stochastic (infinite degree of freedom) time series with a $1/f^\alpha$ power spectrum can exhibit a finite correlation dimension, even for arbitrarily large data sets. [A. R. Osborne and A. Provenzale, *Physica D* **35**, 357 (1989).] I will discuss the relevance of this observation to the practical estimation of dimension from a time series, and in particular I will argue that a good dimension algorithm need not be trapped by this anomalous fractal scaling. Further, I will analytically treat the case of gaussian $1/f^\alpha$ noise, with explicit high and low frequency cutoffs, and derive the scaling of the correlation integral $C(N, r)$ in various regimes of the $(N, r)$ plane.

[1] *E-mail:* `jt@t13.lanl.gov`
*Telephone:* (505) 665-5682

# 1 Introduction

While a number of algorithms for estimating degrees of freedom directly from a time series have been proposed (see Ref. [1] for a review), the correlation dimension of Grassberger and Procaccia [2] remains the most popular. Although these algorithms are known to be fraught with pitfalls [1, 3, 4, 5], it is generally considered that with a long enough time series, it should be possible in principle for a numerical algorithm to distinguish low-dimensional chaos from stochastic noise. This notion was recently challenged in a remarkable paper by Osborne and Provenzale [6], hereafter OP, which found that for a stochastic time series with a $1/f^\alpha$ power spectrum, the numerical estimate of correlation dimension is a small finite value $D = 2/(\alpha - 1)$ for $1 < \alpha < 3$, even for arbitrarily long time series.

This observation raises at least two serious questions. One, can such an algorithm be fooled into incorrectly characterizing a stochastic time series as low-dimensional? Two, is it possible or even common for low-dimensional chaos to exhibit a $1/f^\alpha$ power spectrum? Sections 2 and 3 will address these questions.

In Section 4, I will discuss the anomalous scaling of OP in qualitative terms. Section 5 will introduce the model of $1/f^\alpha$ noise with high and low frequency cutoffs; section 6 will develop some notation and useful intermediate results; and section 7, the correlation integral $C(N, r)$ will be derived, and several distinct scaling regimes in the $(N, r)$ plane will be delineated. Thus, the practical limitations of real-world ("almost $1/f^\alpha$") noise on the correlation dimension estimator can be quantified.

# 2 Fooling a dimension estimator

The suggestion of OP is that numerical estimation of dimension may not be adequate to the task of distinguishing low-dimensional chaotic systems from stochastic processes. This section will address the question "Can a good dimension algorithm be fooled?"

## 2.1 Stationarity

The first requirement of a time series which is to be analyzed by a dimension algorithm is that it be stationary. No sensible researcher would attempt to estimate the dimension of (say) the Lorenz attractor from a time series that is only half a cycle long, no matter how many points are in the time series. Blind application of a dimension algorithm might give the answer that the dimension is one, and of course it is true that such a short segment of the trajectory is one-dimensional. But the application of common sense would reject this answer as irrelevant to the geometry of the strange attractor, of which the small segment is a poor sample. In general, it is unwise to attempt to compute the dimension of a strange attractor unless the length of the data set is much longer than the length of the dominant period. This point has also made by Mayer-Kress [7], who notes that for $1/f^\alpha$ noise, the dominant period is always on the order of the length of the time series. The operational definition of stationarity is that there be no significant power at low frequencies, and this is violated by $1/f^\alpha$ noise.

## 2.2 Autocorrelation

An important property of $1/f^\alpha$ noise is that the autocorrelation time is long. The sensitivity of dimension algorithms to autocorrelation is well known [1, 4, 8]. For finite autocorrelation, one can be assured that a dimension algorithm will not give spuriously low dimension only if the length of the time series is much longer (sometimes *very* much longer) than the autocorrelation time. If, as in the case of $1/f^\alpha$ noise, the autocorrelation time scales with the size of the data set, then it is no longer sufficient to say that the $N \to \infty$ limit will fix the problem.

In fact, a good estimator of correlation dimension will avoid all pairs of points closer together in time than the autocorrelation [8]. For $1/f^\alpha$ noise, this excludes *all* the pairs; the output of the algorithm will not be a spurious dimension, but a plea for more data.

## 2.3 The method of surrogate data

In any case, it is recommended that the effects of linear correlation be guarded against with a test such as the following. Given a time series, create an ensemble (in practice, five or ten) of "surrogate" time series which are random except that they have the same variance and autocorrelation as the original time series. A practical way to do this is to take the Fourier transform of the original data, randomize the phases, and then take the inverse Fourier transform. Now, any statistic computed from the original data (dimension, Lyapunov exponent, *etc.*) should then be computed from the surrogate data as well. If the results of the computation are significantly different for the original data compared to the surrogate data (and since there are several sets of surrogate data, the statistical significance can be quantified), then the null hypothesis that the the original time series is linearly correlated noise can be rejected.

Attempting to reject the null hypothesis is a less ambitious goal than attempting to estimate the dimension, but if the null hypothesis cannot be rejected, then there is no point estimating dimension. Applied to a time series generated by Fourier inverting a $1/f^\alpha$ spectrum (this is how $1/f^\alpha$ noise is generated in OP), this test will by construction fail to reject the null hypothesis.

A simplified version of this approach was used by Grassberger [9] to reject a claim of low-dimensionality in a climatic time series. The systematic approach advocated here is also described in Refs. [1, 4].

## 3 The power spectrum of a chaotic attractor

The ubiquity of $1/f$ noise sources in physical systems is well established [10, 11, 12], and remains a subject of much interest. Recently, a variety of nonlinear dynamical systems have been investigated which exhibit power spectra with $1/f^\alpha$ scaling.

Pomeau and Manneville [13] observed that intermittent behavior near a critical bifurcation point could exhibit a $1/f^\alpha$ spectrum; this was quantified with a renormalization-group by Procaccia and Schuster [14]. Here, the $1/f^\alpha$ behavior is only seen at non-generic critical points. Also in this case, the motion is bounded, which requires $\alpha \leq 1$.

Keeler and Farmer [15] observed $1/f^\alpha$ spectra with $\alpha$ near 1 in a spatio-temporal system of coupled maps. The "sandpile" model of Bak, Tang, and Weisenfeld [16], also exhibits $1/f^\alpha$ spectra, as part of its self-organized critical behavior. In both of these cases, the dynamical system has many degrees of freedom. Further, there is a low frequency cutoff which corresponds to the finite spatial size of the system. A simple Hamiltonian model of a particle in an unbounded periodic potential (*e.g.*, the motion of a classical electron in a crystal) was described by Geisel, *et al.* [17]. With only four degrees of freedom, the dynamics of this system exhibits a $1/f^\alpha$ spectrum, with $\alpha > 1$, and with no apparent cutoff at low frequency (though, presumably, if the crystal was of limited size, then the motion would be bounded, and a low frequency cutoff would be observed.)

Osborne and Caponio [18] contrived a Hamiltonian with a power-law scaling, and the resulting dynamics were then shown to scale as $1/f^\alpha$ over a limited range of frequencies.

In the above examples, the $1/f^\alpha$ scaling at low frequencies was emphasized. It is also the case that for a dynamical system that arises from a set of coupled infinitely differentiable ordinary differential equations, the power at high frequency must decrease faster than any power-law. The high frequency scaling for noisy dynamical systems is a topic of significant interest (see Refs. [19, 20, 21] for discussion from various points of view), but for the purposes of this Letter, I will argue in a subsequent section that it is the low frequency scaling that is important.

While a true $1/f^\alpha$ spectrum cannot be generated by chaotic motion on a bounded low-dimensional strange attractor, power spectra which exhibit $1/f^\alpha$ scaling over a wide range of frequencies are quite common in practice. From this point of view, the results of OP are indeed relevant to the practical estimation of dimension from a time series that might arise from low-dimensional chaos.

The practical solution is the same as has been described earlier for ordinary autocorrelated data [8]: take more data than the characteristic autocorrelation time $\tau$, and avoid pairs of points closer together in time than $\tau$. Still, it is interesting to see in more detail what effect an almost $1/f^\alpha$ power spectrum has on the numerically estimated correlation dimension.

## 4 Qualitative discussion in terms of crinkliness

The concept of fractal dimension can be applied to time series in two quite distinct ways: one, to indicate the number of degrees of freedom in the underlying dynamical system; and two, to quantify the self-affinity (or "crinkliness") of the trajectory through phase space. Although it is the first which is of interest here, the anomalous scaling observed in OP can be understood in terms of the second. Indeed, Osborne and Provezale explain the low dimension in terms of the fractal self-affine paths through phase space that are generated by the $1/f^\alpha$ spectrum. While crinkliness is (at least intuitively) a high frequency phenomenon, I will argue that it is really the the low frequency or long time correlation that is important.

**Fractional Brownian motion**

An early association of fractal dimension with power-law spectra is cited by Mandelbrot and Van Ness [22] in their discussion of fractional Brownian motion. Here, an exponent $H$, with

$0 \leq H \leq 1$, describes the power-law "diffusion" rate:

$$\langle (x_t - x_{t-T})^2 \rangle^{1/2} \sim T^H \tag{1}$$

Ordinary diffusion corresponds to $H = \frac{1}{2}$. The motion is necessarily crinkly if $0 < H < 1$ for small times $T$, while smooth motion leads to $H = 1$ for small $T$. A long term linear drift gives $H = 1$ for large T, and bounded motion implies $H = 0$ in the large $T$ limit. Fractional Brownian motion exhibits a $1/f^\alpha$ power spectrum with $\alpha = 1 + 2H$.

**Graph dimension**

One way to quantify crinkliness is in terms of the graph dimension, which is defined as the fractal dimension of the set of points $(t, x_t) \in \mathbf{R}^2$. The graph dimension for fractional Brownian motion is given by $D_g = 2 - H$ [23]. For $1/f^\alpha$ noise, since $H = (\alpha - 1)/2$, we have $D_g = (5 - \alpha)/2$ for $1 < \alpha < 3$. Note that $D_g$ saturates at one (that is, the graph loses its crinkliness) for $\alpha \geq 3$.

**Dimension of a local segment**

Alternatively, one can define a fractal dimension for a segment of the trajectory itself. Let $x_t^{(1)}, \ldots, x_t^{(m)}$ denote $m$ components of a multidimensional time series, and consider the trajectory of these $m$ components in $\mathbf{R}^m$. Over a small interval of time $T$, the distance between endpoints of the segments is equal to $r = \max_i |x_t^{(i)} - x_{t-T}^{(i)}|$. The local dimension $D$ is given by the scaling of "bulk" (number of points, $T$) with "size" ($r$); that is, $T \sim r^D$. For fractional Brownian motion, $\langle (x_t - x_{t-T})^2 \rangle^{1/2} \sim T^H$ implies a dimension $D = 1/H$. Thus, for $1/f^\alpha$ noise, the dimension is given by $D = 2/(\alpha - 1)$ for $1 < \alpha < 3$, and $D = 1$ for $\alpha \geq 3$.

**Recurrence time**

The dimension for a local segment of a fractal path is the same as the dimension of the full trajectory (in the $N \to \infty$ limit) only if the trajectory is non-recurrent. For a recurrent trajectory, the total number of points (the "bulk") within $r$ of a fiducial point will be much greater than the $T$ above, because a recurrent trajectory keeps coming back to the fiducial point.

More crucial than the high frequency crinkles, then, is the relationship between the low frequency cutoff and the length of time series; and whether or not the trajectory through phase space is recurrent. If there is an explicit low frequency cutoff and an arbitrarily long time series, then eventually the trajectory will arbitrarily many times come back arbitrarily close to a fiducial point on the trajectory. On the other hand, if the length of the time series is limited, and the cutoff $f_o$ is very small, then the trajectory may be effectively non-recurrent.

For the non-recurrent case, the dimension of the full trajectory will be equal to the dimension of a local segment. But if the time series is long enough to be recurrent, the phase space will be "filled out" and the dimension estimated will be that of the embedding space. One property of $1/f^\alpha$ noise (without cutoffs) is that it is not recurrent for $m > 2/(\alpha - 1)$ where $m$ is the embedding dimension. (As an example, an ordinary random walk exhibits a $1/f^2$ spectrum, and is known to be recurrent only in one and two dimensions.

In three dimensions, there is a finite probability that the trajectory will never return to a neighborhood of the starting point.) From this point of view, it is not surprising that the dimension of "pure" $1/f^\alpha$ noise should saturate at $D = 2/(\alpha - 1)$ for large $m$.

**Finite dimension in non-power-law spectra**

Inspired by the results of OP, some researchers have looked for finite fractal dimension in time series with spectra that are not strictly power-law. For instance, Greis and Greenside [24] observe fractal scaling in a class of stochastic Langevin equations. This class includes Ornstein-Uhlenbeck noise, the power spectrum of which is approximately constant for low frequency, and $1/f^2$ for high frequency. For small $N$, an apparent dimension of $D = 2$ is observed; for larger $N$ however, the dimension approaches the embedding dimension. It is important to distinguish the scaling these authors sometimes seek (which is a measure of crinkliness) from what correlation dimension attempts to quantify (which is number of degrees of freedom). In fact, correlation dimension is probably not a very good measure of crinkliness, compared to the graph dimension (for which recurrence is explicitly prohibited) or a direct measure of the diffusion exponent $H$. See Refs. [25, 26, 27] for discussion of numerical issues in this context.

## 5 Explicit and implicit frequency cutoffs

It is important to note that a pure $1/f^\alpha$ power spectrum is physically impossible. Depending on whether $\alpha$ is greater or less than one, there will be an infrared or ultraviolet catastrophe – that is, the total power will diverge to infinity. There is therefore a need to introduce frequency cutoffs in order to analyze the properties of $1/f^\alpha$ noise.

Even without explicit frequency cutoffs, there are implicit cutoffs for real time series. The low frequency is effectively cut off by the finite length of the time series, and the high frequency is cut off by the sampling rate.

In what follows a particular stochastic time series will be discussed. A $1/f^\alpha$ spectral shape will be considered, with a cutoff at low freqency $f_o$ and at high frequency $f_1$. (I assume $f_1 \gg f_o$.) In the absence of explicit cutoffs, I will take $f_o \approx 1/N$, and $f_1 \approx 1$. From this model, the correlation integral $C(N, r)$ will be derived analytically for various regimes of $N$ and $r$, as a function of the cutoffs $f_o$ and $f_1$, the exponent $\alpha$, and the embedding dimension $m$ (which, in general, will be taken to be large).

The particular power spectrum that will be considered here exhibits $1/f^\alpha$ scaling between a lower and upper cutoff frequency. Below the lower cutoff, the spectrum is flat, and above the upper cutoff, the spectrum is zero.

$$S(f) = \begin{cases} f_o^{-\alpha} & \text{for } f \leq f_o \\ f^{-\alpha} & \text{for } f_o \leq f \leq f_1 \\ 0 & \text{for } f_1 < f \end{cases} \quad (2)$$

These cutoffs ensure that the time series has a well defined variance:

$$\langle x^2 \rangle = \int_0^\infty S(f) df = f_o^{1-\alpha} + (f_o^{1-\alpha} - f_1^{1-\alpha})/(\alpha - 1). \quad (3)$$

For $\alpha \leq 1$, the high frequency cutoff is necessary to prevent an ultraviolet catastrophe, and for $\alpha \geq 1$, the low cutoff is necessary to prevent divergence at low frequencies. Since anomalous estimated dimension occurs for $\alpha > 1$, that is the case that will be considered here. Then, we can write

$$x_{\text{rms}}^2 := \langle x^2 \rangle \approx \frac{\alpha}{\alpha - 1} f_o^{1-\alpha}. \tag{4}$$

The autocorrelation function can be expressed as a Fourier transform of the power spectrum:

$$\begin{aligned}
A(T) &:= \frac{1}{x_{\text{rms}}^2} \int_0^\infty \cos(2\pi f T) S(f) \, df \\
&= \frac{1}{x_{\text{rms}}^2} \left[ \int_0^{f_o} \cos(2\pi f T) f_o^{-\alpha} \, df + \int_{f_o}^{f_1} \cos(2\pi f T) f^{-\alpha} \, df \right]
\end{aligned} \tag{5}$$

From this, the variance of the differenced time series can also be calculated; here

$$\begin{aligned}
\sigma_T^2 &:= \langle (x_t - x_{t-T})^2 \rangle = 2 x_{\text{rms}}^2 - 2 \langle x_t x_{t-T} \rangle = 2 x_{\text{rms}}^2 (1 - A(T)) \tag{6} \\
&= \int_0^\infty 4 \sin^2(\pi f T) S(f) \, df. \tag{7}
\end{aligned}$$

The behavior of this statistic at various ranges of $T$ will prove to be important for subsequent analysis, and so this will be derived in some detail in the next section.

# 6 Behavior of the statistic $\sigma_T^2 := \langle (x_t - x_{t-T})^2 \rangle$

In this section, I will calculate the statistic $\sigma_T^2$ for the example of truncated $1/f^\alpha$ noise described above. This approach is just the reverse of Mandelbrot and Van Ness [22], who define fractional Brownian motion by requiring *a priori* that $\sigma_T \sim T^H$ for some exponent $H$.

From Eq. (7), we can write

$$\begin{aligned}
\sigma_T^2 &= \int_0^{f_o} 4 \sin^2(\pi f T) f_o^{-\alpha} \, df + \int_{f_o}^{f_1} 4 \sin^2(\pi f T) f^{-\alpha} \, df \\
&= 2 f_o^{1-\alpha} \left[ 1 - \frac{\sin(2\pi f_o T)}{2\pi f_o T} \right] + \int_{f_o}^{f_1} 4 \sin^2(\pi f T) f^{-\alpha} \, df.
\end{aligned} \tag{8}$$

The asymptotic behavior of this integral expression for $\sigma_T^2$ can be evaluated in each of three regimes. We find $\sigma_T \sim T$ for very short time $T$, $\sigma_T \sim T^{(\alpha-1)/2}$ for $T$ in an intermediate regime, and $\sigma_T \sim$ constant for very large $T$. Fig. 1 shows how $\sigma_T$ varies with $T$ in each of these three regimes.

Figure 1: Variation with $T$ of the statistic $\sigma_T^2 = \langle (x_t - x_{t-T})^2 \rangle$

## 6.1 I: Very short time, $T \ll 1/f_1$

Here, $\sin \pi fT \approx \pi fT$, and $1 - \sin(2\pi f_o T)/(2\pi f_o T) \approx \frac{1}{6}(2\pi f_o T)^2$, so Eq. (8) can be approximated

$$\begin{aligned}
\sigma_T^2 &\approx 2f_o^{1-\alpha}\left[\frac{1}{6}(2\pi f_o T)^2\right] + 4\pi^2 T^2 \int_{f_o}^{f_1} f^{2-\alpha}\,df \\
&= (2\pi T)^2 \left(f_1^{3-\alpha} - \frac{\alpha}{3}f_o^{3-\alpha}\right)/(3-\alpha) \\
&= \begin{cases} \frac{4\pi^2}{3-\alpha}f_1^{3-\alpha}T^2 & \text{for } \alpha < 3 \\ \frac{4\pi^2\alpha}{3(3-\alpha)}f_o^{3-\alpha}T^2 & \text{for } \alpha > 3. \end{cases}
\end{aligned} \qquad (9)$$

Thus, $\sigma_T \sim T$ for small $T$, which implies no crinkles; this arises as a direct consequence of the high frequency cutoff.

## 6.2 II: Intermediate time, $1/f_1 \ll T \ll 1/f_o$

Now, let $u = fT$, so

$$\begin{aligned}
\sigma_T^2 &\approx \frac{1}{3}f_o^{3-\alpha}(2\pi T)^2 + 4T^{\alpha-1}\left[\int_{f_o T}^{f_1 T} \sin^2(\pi u)\, u^{-\alpha}\, du\right] \\
&\approx 4I(\alpha)T^{\alpha-1}
\end{aligned} \qquad (10)$$

where the multiplier

$$I(\alpha) := \int_0^\infty \sin^2(\pi u)\, u^{-\alpha}\, du \qquad (11)$$

converges to a constant for $1 < \alpha < 3$, and is of order unity as long as $\alpha$ is not too close to 1 or to 3. (Asymptotically, $I(\alpha) \sim \frac{1}{2(\alpha-1)}$ for $\alpha \to 1$; and $I(\alpha) \sim \frac{\pi^2}{3-\alpha}$ for $\alpha \to 3$.)

Figure 2: Integral $I(\alpha) = \int_0^\infty \sin^2(\pi u)\, u^{-\alpha}\, du$

The scaling in this intermediate regime corresponds to the scaling postulated for fractional Brownian motion, with $H = (\alpha - 1)/2$.

For $\alpha > 3$, the integral in Eq. (11) diverges, and it is not justified to approximate the lower limit $f_o T$ as zero. In this case, we obtain that the leading order is equal to the small time limit given in Eq. (9). In this case, there is no distinction between regimes I and II. Intuitively, $\alpha > 3$ corresponds to not having enough high frequency to "crinkle" the path.

## 6.3 III: Very long time, $1/f_o \ll T$

In this case, let $u = f/f_o$, so

$$\begin{aligned}
\sigma_T^2 &= 2f_o^{1-\alpha}\left[1 - \frac{\sin(2\pi f_o T)}{2\pi f_o T}\right] + 4f_o^{1-\alpha}\int_1^{f_1/f_o} \sin^2(\pi f_o T u)\, u^{-\alpha}\, du \\
&\approx 2f_o^{1-\alpha}\left[1 - \frac{\sin(2\pi f_o T)}{2\pi f_o T}\right] + 2f_o^{1-\alpha}\int_1^\infty [1 - \cos(2\pi f_o T u)]\, u^{-\alpha}\, du
\end{aligned}$$

$$\approx 2f_o{}^{1-\alpha}\left[1-\frac{\sin(2\pi f_o T)}{2\pi f_o T}\right]+2f_o{}^{1-\alpha}\left[\frac{1}{(\alpha-1)}+\frac{\sin(2\pi f_o T)}{2\pi f_o T}-\alpha\frac{\cos(2\pi f_o T)}{(2\pi f_o T)^2}\right]$$

$$= 2f_o{}^{1-\alpha}\left[\frac{\alpha}{\alpha-1}-\alpha\frac{\cos(2\pi f_o T)}{(2\pi f_o T)^2}\right] \tag{12}$$

But, $x_{\text{rms}}^2 = \alpha f_o{}^{1-\alpha}/(\alpha-1)$, so we can write

$$\sigma_T^2 \approx 2x_{\text{rms}}^2\left[1-(\alpha-1)\frac{\cos(2\pi f_o T)}{(2\pi f_o T)^2}\right]. \tag{13}$$

### 6.4 Crossover times

The crossover times between these behaviors occurs roughly at $T_o \sim 1/f_1$ and $T_1 \sim 1/f_o$. However, it is possible to more precisely specify these times by equating the formulas for $\sigma_T^2$ in the different regimes.

For the first crossover time, assume $1 < \alpha < 3$ (otherwise, there is no crossover behavior), and write $(2\pi T)^2 f_1^{3-\alpha}/(3-\alpha) = 4I(\alpha)T^{\alpha-1}$, which, upon solving for $T$, gives the crossover time

$$T_o = \left[\frac{(3-\alpha)I(\alpha)}{\pi^2}\right]^{1/(3-\alpha)}(1/f_1) \tag{14}$$

For the second crossover time, write $4I(\alpha)T^{\alpha-1} = 2f_o{}^{1-\alpha}/(\alpha-1)$ to obtain

$$T_1 = [2(\alpha-1)I(\alpha)/\alpha]^{1/(1-\alpha)}(1/f_o) \tag{15}$$

These coefficents of the $1/f_1$ and $1/f_o$ are "only" numerical factors, but they are still useful for more precisely pinning down the scaling range.[2]

In particular, note that the range for intermediate (fractal) scaling of $\sigma_T$ with $T$ is given by

$$\frac{T_1}{T_o} = K(\alpha)\frac{f_1}{f_o} \tag{16}$$

where the numerical factor is given by

$$K(\alpha) = \frac{[2(\alpha-1)I(\alpha)/\alpha]^{1/(1-\alpha)}}{[(3-\alpha)I(\alpha)/\pi^2]^{1/(3-\alpha)}}. \tag{17}$$

For $\alpha$ near 1 or 3, this factor can be extremely small, which implies that the range over which anomalous fractal scaling is observed can also be very small. This also partly explains the difficulty that both Fox [26] and Higuchi [27] describe in getting numerical agreement with theory for $\alpha$ near 1 and 3.

---

[2]For example, there is an unexplained factor of approximately $\frac{1}{5}$ that appears in Figure 3 of Higuchi's paper [25]. His graph dimension of 1.21 corresponds to $\alpha = 2.58$, which gives $f_o \approx 0.15/T_o$, which is consistent with the figure.

### 6.4.1 Implicit frequency cutoffs

In the absence of explicit cutoffs, the effective frequency cutoffs are defined by crossover times of $T_o = 1$ and $T_1 = N$. From Eqs. (14, 15), this gives

$$f_1 = \left[(3-\alpha)I(\alpha)/\pi^2\right]^{1/(3-\alpha)} \tag{18}$$

$$f_o = \left[2(\alpha-1)I(\alpha)/\alpha\right]^{1/(1-\alpha)}/N \tag{19}$$

## 6.5 Characteristic length scales

Associated with the characteristic times $T = 1$, $T = T_o$, and $T = T_1$ are characteristic distance scales. Define

$$R_o = \sqrt{2}\sigma_{T=1} = \begin{cases} 2\pi\sqrt{\frac{2}{3-\alpha}}f_1^{(3-\alpha)/2} & \text{for } \alpha < 3 \\ 2\pi\sqrt{\frac{2\alpha}{3(\alpha-3)}}f_o^{(3-\alpha)/2} & \text{for } \alpha > 3. \end{cases} \tag{20}$$

and

$$R_1 = \sqrt{2}\sigma_{T_o} = [8I(\alpha)]^{1/2}T_o^{(\alpha-1)/2}. \tag{21}$$

In between these two distances a third will be defined

$$R_* = [8I(\alpha)]^{1/2}. \tag{22}$$

Note that in the case $T_o = 1$ (no explicit high frequency cutoff), we have $R_* = R_1 = R_o$. Finally, the largest distance scale is $x_{\text{rms}}$ which is also equal to $\sqrt{2}\sigma_{T_1}$. An expression for this appears in Eq. (4).

# 7 Derivation of the correlation integral

Using the results and definitions of the previous section, I will derive an analytical expression for the expected value of the correlation integral $C(N, r)$ in the case of $1/f^\alpha$ noise with cutoffs, and then show how the expression scales in various regimes of $N$ and $r$.

Consider $P_T(r)$ as the probability that a pair of points $(x_{t-T}, x_t)$ satisfies $(x_t - x_{t-T})^2 < r^2$. For the gaussian noise considered here, $P_T(r) = \text{erf}(\frac{r}{\sqrt{2}\sigma_T})$. If the trajectory is embedded in $m$ dimensions, this becomes $P_T(r) = \text{erf}(\frac{r}{\sqrt{2}\sigma_T})^m$, assuming the $L_\infty$ or "max" norm. This exact form relies on the assumption that each component of the embedded time series is independent. In OP, this is ensured by taking separate realizations of $1/f^\alpha$ noise for each coordinate. Although a delay-time embedding does not strictly satisfy this criterion, it has been observed numerically that the same qualitative results are obtained [6, 8].

Now the correlation integral is given by $C(N, r) = (2/N^2)\sum_T (N-T)P_T(r)$. Thus,

$$C(N, r) = \frac{2}{N^2}\sum_{T=1}^{N}(N-T)\text{erf}(\frac{r}{\sqrt{2}\sigma_T})^m \tag{23}$$

is an exact expression for $C(N, r)$ in terms of the statistic $\sigma_T$.

It is important to recognize that the function erf has as asymptotic behavior, $\text{erf}(x) \approx x$ for small $x$, and $\text{erf}(x) \approx 1$ for large $x$. Thus the sum can be broken into two components, one for which $r < \sqrt{2}\sigma_T$ and one for which $r > \sqrt{2}\sigma_T$. Let $T(r)$ define where this transition occurs:

$$\sqrt{2}\sigma_{T(r)} = r. \tag{24}$$

Then,

$$\begin{aligned}
C(N,r) &= \frac{2}{N^2}\left[\sum_{T=1}^{T(r)}(N-T)\text{erf}(\frac{r}{\sqrt{2}\sigma_T})^m + \sum_{T=T(r)}^{N}(N-T)\text{erf}(\frac{r}{\sqrt{2}\sigma_T})^m\right] \\
&\approx \frac{2}{N^2}\left[\sum_{T=1}^{T(r)}(N-T) + \sum_{T=T(r)}^{N}(N-T)(\frac{r}{\sqrt{2}\sigma_T})^m\right] \\
&= \frac{2}{N^2}\left[(NT(r) - \frac{1}{2}T(r)^2) + r^m 2^{-m/2}\sum_{T=T(r)}^{N}(N-T)\sigma_T^{-m}\right] \\
&= \frac{2T(r)}{N}\left(1 - \frac{T(r)}{2N}\right) + \frac{2^{1-m/2}}{N^2}r^m\left[\sum_{T=T(r)}^{N}N\sigma_T^{-m} - \sum_{T=T(r)}^{N}T\sigma_T^{-m}\right] \tag{25}
\end{aligned}$$

This expression, though a bit unwieldy as it stands, will be shown to exhibit a variety of different scalings in different regimes, beginning with the fractal scaling observed in OP.

## 7.1 Regime of fractal scaling

Assume that $r$ is sufficiently large, and $N$ is sufficiently small that the scaling $\sigma_T \sim T^{(\alpha-1)/2}$ holds for all $T(r) \leq T \leq N$. This requires $\alpha < 3$, and $N \leq T_1 \approx 1/f_o$ and $T(r) \geq T_o \approx 1/f_1$, or $r \geq \sqrt{2}\sigma_{T_o} = R_1$. This argument will also require $m > 4/(\alpha - 1)$. Now,

$$\begin{aligned}
\sum_{T=T(r)}^{N} N\sigma_T^{-m} &\approx \int_{T(r)}^{N} N\sigma_T^{-m} \approx N[4I(\alpha)]^{-m/2}\int_{T(r)}^{N}T^{-m(\alpha-1)/2} \\
&= \frac{N[4I(\alpha)]^{-m/2}}{1 - m(\alpha-1)/2}\left[T(r)^{1-m(\alpha-1)/2} - N^{1-m(\alpha-1)/2}\right] \\
&= \frac{N[4I(\alpha)]^{-m/2}}{1 - m(\alpha-1)/2}T(r)^{1-m(\alpha-1)/2}\left[1 - \left(\frac{T(r)}{N}\right)^{m(\alpha-1)/2-1}\right] \tag{26}
\end{aligned}$$

The second term is negligible if $N \gg T(r)$ and $m > 2/(\alpha - 1)$, and in this case

$$\sum_{T=T(r)}^{N} N\sigma_T^{-m} \approx \frac{N[4I(\alpha)]^{-m/2}}{1 - m(\alpha-1)/2}T(r)^{1-m(\alpha-1)/2} \tag{27}$$

Similarly,

$$\sum_{T=T(r)}^{N} T\sigma_T^{-m} \approx \frac{[4I(\alpha)]^{-m/2}}{2 - m(\alpha-1)/2}T(r)^{2-m(\alpha-1)/2} \tag{28}$$

but this time, the condition for neglecting the extra term is that $m > 4/(\alpha - 1)$. Also, note that since $N \gg T(r)$, we can neglect the $\sum_{T=T(r)}^{N} T\sigma_T^{-m}$ term because it is much less than $\sum_{T=T(r)}^{N} N\sigma_T^{-m}$.

For convenience, write

$$D := 2/(\alpha - 1) \tag{29}$$

Now, use $\sigma_{T(r)} \approx [4I(\alpha)]^{1/2} T(r)^{(\alpha-1)/2}$ (from Eq. (10), which is valid in this regime), with Eq. (24) which defines $T(r)$, and Eq. (22) which defines $R_*$, to obtain

$$T(r) = \left(\frac{r}{R_*}\right)^D. \tag{30}$$

Then from Eq. (25), we have

$$\begin{aligned} C(N, r) &\approx \frac{2}{N}\left(\frac{r}{R_*}\right)^D + \frac{2^{1-m/2}}{N^2} r^m \left[\frac{N(R_*/\sqrt{2})^{-m}}{1 - m/D} T(r)^{1-m/D}\right] \\ &= \frac{2}{N}\left[\frac{m - 2D}{m - D}\right]\left(\frac{r}{R_*}\right)^D. \end{aligned} \tag{31}$$

This demonstrates the $C(N, r) \sim r^D$ scaling that was observed in OP. Apart from providing an alternative derivation of the scaling that is applicable directly to the correlation integral, it gives a sense of how observable the phenomena might be. For instance, it shows that $\lim_{N \to \infty} C(N, r) = 0$. For sufficently large $N$, as will be shown in the next section, the contribution of the fractal term becomes negligible, and $C(N, r) \sim r^m$ scaling will dominate.

The $m$ dependence of the expression above for $C(N, r)$ is also interesting. (First of all, recall that the expression was derived under the assumption $m > 2D$ and so is valid only for large $m$.) Grassberger and Procaccia [28] define an entropy $K_2$ which expresses the scaling of the correlation integral with $m$.

$$C_m(N, r) \sim e^{-mK_2} r^\nu \tag{32}$$

This kind of exponential dependence on $m$ is not seen in Eq. (31). In contradiction to what is claimed in Ref. [6], I find that the correlation entropy does not have a "finite and predictable value", but instead gives $K_2 = 0$ in the limit $m \to \infty$.

## 7.2 Very large $N$

For $N$ sufficiently large, I can show that $C(N, r)$ will not exhibit the fractal scaling observed in OP; but instead will scale as $C(N, r) \sim r^m$ indicating a stochastic system. First, break Eq. (23) into two terms:

$$C(N, r) = \frac{2}{N^2} \sum_{T=1}^{T_1} (N - T)\text{erf}(\frac{r}{\sqrt{2}\sigma_T})^m + \frac{2}{N^2} \sum_{T=T_1}^{N} (N - T)\text{erf}(\frac{r}{\sqrt{2}\sigma_T})^m \tag{33}$$

The first term was calculated in the previous section on fractal scaling. To calculate the second term, note that $\sigma_T \approx \sqrt{2} x_{\text{rms}}$ for all $T > T_1$, so for $r < 2x_{\text{rms}}$, we have

$$C(N, r) = \frac{2}{N}\left[\frac{m - 2D}{m - D}\right]\left(\frac{r}{R_*}\right)^D + (1 - 2T_1/N)\left(\frac{r}{2x_{\text{rms}}}\right)^m \tag{34}$$

This sum will be dominated by the second term as long as

$$N \gg 2 \left(\frac{r}{R_*}\right)^D \left(\frac{r}{2x_{\text{rms}}}\right)^{-m} \sim r^{D-m}. \quad (35)$$

This implies that the number of needed data points $N$ increases for smaller $r$. If this condition holds for all $r > R_*$, then

$$N \gg 2 \left(\frac{2x_{\text{rms}}}{R_*}\right)^m \approx 2 \left(\frac{\alpha}{2(\alpha-1)I(\alpha)}\right)^{m/2} f_o^{m(1-\alpha)/2} \quad (36)$$

in which case

$$C(N,r) = \left(\frac{r}{2x_{\text{rms}}}\right)^m \quad (37)$$

Note that in this case, unlike the case of the fractal scaling regime, the limit $\lim_{N\to\infty} C(N,r)$ is nonzero. It is this scaling that ultimately dominates, although as Eq. (36) shows, $N$ may have to be extremely large before this happens.

As this argument shows, it is the low frequency cutoff $f_o$ that is crucial. Another way to think of this is in terms of the average recurrence time $T(\epsilon)$ for a trajectory to come back to within $\epsilon$ [7]. In this case,

$$T(\epsilon) = \lim_{N\to\infty} 1/C(N,\epsilon) = (2x_{\text{rms}}/\epsilon)^m = \frac{2^m[\alpha/(\alpha-1)]^{m/2}}{\epsilon^m} f_o^{m(1-\alpha)/2} \quad (38)$$

If $N \gg T(\epsilon)$, then good scaling is expected for $C(N,r)$ for $r > \epsilon$. Note that for large embedding dimension $m$, the data requirement is extreme; so that even to say $N \gg 1/f_o$ does not guarantee that spurious scaling will not be observed. In the case of Ornstein-Uhlenbeck noise, $\alpha = 2$ and $\tau = 1/f_o$, so that this condition becomes $N \gg \tau^{m/2}$, in agreement with one of the main results of Ref. [8].

### 7.3 Effect of high frequency cutoff $f_1$

While a low frequency cutoff is essential to prevent an infrared catastrophe, the high frequency cutoff is optional for $\alpha > 1$. When there is a high frequency cutoff, the trajectory loses its crinkly shape at small distances, and locally looks one dimensional. This can be reflected in the correlation integral as a scaling of $C(N,r) \sim r$ over a range of $r$.

In the range $T < T_o$, we have $\sigma_T = \sigma_{T=1} T$ and Eq. (9) gives an expression for $\sigma_{T=1}$. Combining this with Eq. (20) for $R_o$ and Eq. (25), we can write

$$C(N,r) = \frac{2}{N}\left(\frac{r}{R_o}\right). \quad (39)$$

This scaling is exhibited in the range $R_o < r < R_1$.

### 7.4 $\alpha > 3$

If $\alpha > 3$, there is not enough high frequency to crinkle the trajectory. The smooth trajectory is locally one-dimensional, and a scaling $C(N,r) \sim r$ is observed. In particular,

$$C(N,r) = \frac{2}{N}\left(\frac{r}{R_o}\right) \quad (40)$$

valid for the range $R_o < r < x_{\text{rms}}$.

## 7.5 Very small $r$

For $r$ sufficiently small that $r < \sqrt{2}\sigma_T$ for all $T$ (that is, for $r < R_o = \sqrt{2}\sigma_{T=1}$), it is possible to approximate $\text{erf}(r/\sqrt{2}\sigma_T)$ by $r/\sqrt{2}\sigma_T$. In this case, Eq. (25) becomes

$$
\begin{aligned}
C(N,r) &= \frac{2^{1-m/2}}{N^2}\left[\sum_{T=1}^{N} N\sigma_T^{-m} - \sum_{T=1}^{N} T\sigma_T^{-m}\right] r^m \\
&= \frac{2}{N}\left[\sum_{T=1}^{N}(1-T/N)(\sigma_T/\sigma_{T=1})^{-m}\right]\left(\frac{r}{R_o}\right)^m
\end{aligned}
\quad (41)
$$

The term in the square brackets is a constant, and in the limit $m \to \infty$ approaches unity, so

$$C(N,r) \approx \frac{2}{N}\left(\frac{r}{R_o}\right)^m \quad (42)$$

In general, regardless of the cutoffs, we find $C(N,r) \sim r^m$ for small $r$; in particular, the "lower half" of the curve, where $C(N,r) < 2/N$, will exhibit $r^m$ scaling indicating a stochastic time series. Since $C(N,r) \geq 2/N^2$, the range in $r$ over which this scaling is valid is $R_o N^{-1/m} < r < R_o$.

## 8 Conclusions

The conclusions can be presented as a "phase diagram" in the $(N,r)$ plane. As Fig. 3 shows, as many as three different scaling exponents can be seen. For sufficiently large $N$, only the correct $C \sim r^m$ scaling is observed, though $N$ may have to be extremely large for this regime to be achieved. The case considered in OP involves no explicit frequency cutoffs, and can be recovered with $1/f_o \approx N$ and $f_1 \approx 1$. In this case, anomalous scaling of $C \sim r^{2/(\alpha-1)}$ is observed for large $r$ and correct scaling of $C \sim r^m$ is observed for small $r$. If the high frequency cutoff is significant ($f_1 \ll 1$), then a third regime is seen, in which $C \sim r$ scaling is observed.

Figure 3: Phase diagram of scaling in $(N,r)$ plane

Fig. 4 shows a representative correlation integral, computed from a time series of $1/f^\alpha$ noise with high and low frequency cutoffs. Notice that three different scaling regimes are visible.

Figure 4: Correlation integral $C(N,r)$ for $1/f^\alpha$ noise.

I should emphasize that all these calculations were done using the original definition of correlation integral, specified by Grassberger and Procaccia [2]. These anomalous regions are further restricted and can even be eliminated if the modified correlation integral defined in Ref. [8] is used. For practical estimation of correlation dimension, it is this modified version that is preferred.

## Acknowledgments

I am grateful to A. R. Osborne, N. Gershenfeld, J. D. Farmer, and the T-13 Time Series Group for useful comments. I especially want to thank H. Greenside for a vigorous and enlightening e-mail dialogue.

# Figures

**Fig. 1.** *Variation of $\sigma_T$ with $T$.*
This figure shows the the three regimes (I, II, and III) of $T$ as defined by the behavior of the statistic $\sigma_T^2 = \langle (x_t - x_{t-T})^2 \rangle$. Shown are the crossover times $T_o$ and $T_1$. Also shown are the distance scales $R_o$, $R_*$, $R_1$, and $x_{\rm rms}$. Note that the axes are logarithmic. In regime I, $\sigma_T \sim T$; in regime II, $\sigma_T \sim T^{(\alpha-1)/2}$; and in regime III, $\sigma_T$ approaches a constant.

**Fig. 2.** *The integral $I(\alpha) = \int_0^\infty \sin^2(\pi u) \; u^{-\alpha} \, du$.*
A number of results derived in this Letter are expressed in terms of this integral, which cannot be explicitly solved. Although $I(\alpha)$ is often treated as being of order unity, note that it can be quite large for $\alpha$ near 1 or 3.

**Fig. 3.** *Phase diagram of scaling in $(N, r)$ plane.*
This figure shows how $C(N, r)$ scales with $r$ in various regimes of the $(N, r)$ plane for $1/f^\alpha$ noise with cutoffs as defined in Eq. (2). This figure assumes $1 < \alpha < 3$ and that the embedding dimension $m$ is large. The axes are logarithmic.

Since the input is inherently stochastic, the correct scaling is $C \sim r^m$. This scaling is observed in areas (a), (b), and (c). In (a), $N$ is sufficiently large (see Eq. (36)) that correct scaling is seen for all $r$. For small $r < R_o$, in (b), correct scaling is observed for all values of $N$. In the third regime, (c), with both $r$ and $N$ large (according to Eq. (35)).

If there is a high frequency cutoff, $f_1 \ll 1$, then there will be an intermediate regime (d) for which $C \sim r$ scaling is observed. Note that the width of this regime is given by $R_1/R_o \sim 1/f_1$, and that if there is no explicit high frequency cutoff, then $f_1 \approx 1$, and this regime vanishes.

If $r > x_{\rm rms}$, in area (e), then $C(N, r)$ will have saturated to unity, and the trivial scaling $C \sim r^0$ will be observed. Further, there will be no scaling in area (f), where $r$ so small that $C(N, r) < 2/N^2$, since there are no distances that small.

Finally, the regime (g) of large $r$ and small $N$ exhibits the anomalous fractal scaling $C \sim r^{2/(\alpha-1)}$ that was observed in OP. If there is no explicit low frequency cutoff (the case considered in OP), then the effective cutoff is given by $f_o = 1/N$, and this anomalous scaling will be observed even for arbitrarily large $N$.

**Fig. 4.** *Correlation integral for $1/f^\alpha$ noise.*
The correlation integral was computed for embedding dimension $m = 1, 2, \ldots, 15$, using $N = 16384$ points in a time series of $1/f^2$ noise with high frequency cutoff $f_1 = 0.05$ and low frequency cutoff $f_o = 0.00006$ (equivalent to no explicit low frequency cutoff). One observes $C \sim r^m$ for small $r$; $C \sim r$ for a very small range of intermediate values of $r$; and the anomalous scaling $C \sim r^{2/(\alpha-1)} = r^2$ for large $r$.